\documentclass[twocolumn,showpacs,preprintnumbers,amsmath,amssymb,prl]{revtex4}
\usepackage{pifont}
\usepackage{amsmath}
\usepackage{amsfonts}
\usepackage{graphicx}
\usepackage{amssymb}
\usepackage{dcolumn}
\usepackage{bm}
\usepackage{slashed}

\begin{document}

\title{Trident Pair Production in Colliding Bright X-ray Laser Beams}

\author{Huayu Hu$^{\ast}$}
\author{Jie Huang}
\affiliation{Hypervelocity Aerodynamics Institute, China Aerodynamics Research and Development Center, Mianyang 621000, Sichuan, P. R. China}

\begin{abstract}
The magnificent development of strong X-ray lasers motivates the advancement of pair production process studies into higher laser frequency region. In this paper, a resonant electron-positron pair production process with the absorption of two X-ray photons is considered in the impact of an energetic electron at the overlap region of two colliding X-ray laser beams.
Laser-dressed QED method is justified to tackle the complexity of the corresponding multiple Feynman diagrams calculation. The dependence of the production rate as well as the positron energy distribution on the relative angles among the directions of the two laser wave vectors  and the incoming electron momentum is revealed. It is shown that the non-plane wave laser field configuration arouses novel features in the pair production process compared to the plane-wave case.
\end{abstract}



\maketitle
Substantial achievements of strong laser facilities make it possible to study nonlinear QED processes in the laboratory. Especially after the first direct observation of multiphoton electron-positron pair production in the collision of a $\approx50$GeV electron and an optical strong laser beam with intensity $\approx10^{18}$W/cm$^2$ at SLAC (Stanford, California) \cite{SLAC}, tremendous theoretical efforts have been devoted to understanding the various features of the nonlinear and/or nonperturbative trident  process ($\it{particle}+ photons\rightarrow scattered\, particle + e^+e^-$), where the laser field strength seen by the projectile is Lorentz-boosted. Mainly two kinds of theoretical models classified by the particle type are investigated extensively: 1) laser-nucleus collision model, where the nucleus can either be treated as a quantum particle to take into account the nuclear recoil effect \cite{nuclearrecoil} or simplified as a Coulomb potential \cite{nuclearnonrecoil}. Besides basic reaction features, the influence of bound atomic states \cite{bound} and the inclusion of an additional XUV photon \cite{XUV} have been studied as well; 2) laser-electron collision model, which is more computationally demanding due to the nonnegligible electron recoil and exchangeability of the two final electrons, has been developed in complete laser-dressed QED theory \cite{HCC, Ilderton2011}. The total production rates and energy spectra of the produced particles from multi-photon perturbative regime to quasistatic regime are obtained \cite{HCC}, envisaging the prospect of an all-optical tabletop setup for exploring the trident process.

With the development of intense X-ray lasers, their potential in studying the trident process becomes notable.  X-ray output with frequency up to $10$keV and intensity $\sim10^{20}$W/cm$^2$ is the goal of X-ray free electron lasers (XFEL) being improved at DESY \cite{XFELDESY} and SLAC \cite{XFELSLAC}. The employment of strong XFEL pulses can significantly reduce the criteria on the projectile energy in trident pair production experiment as smaller Lorentz factors may be sufficient \cite{Avetissian}, and thus favors the realization of an all-laser setup where the energetic projectile can be provided by a laser-plasma wakefield accelerator, which has achieved the boost of electrons to more than 1GeV \cite{wake}. XFEL-proton collision \cite{xfelpiazza} and XFEL-electron collision \cite{HCC} have been considered for $e^+e^-$ pair production, and high-energy XFEL-nucleus collision is also proposed for $\mu^+\mu^-$ production \cite{xfelmu}.

In this paper besides exploring the performance of high frequency lasers, another aim is to investigate the influence of non-plane-wave laser configuration on the trident pair production process. Here particularly a $50$MeV electron impacting at the overlap region of two colliding X-ray laser beams with frequency $10$keV and intensity $\sim10^{20}$W/cm$^2$, respectively, is assumed. Although the colliding beam configuration is an effective way to raise the field intensity and is demonstrated in some cases to have special effect in enhancing the pair yield \cite{collidingbeams}, it has not been examined before in the trident process as far as we know.

Plane-wave laser field assumption is dominantly resorted to in previous complete QED calculations for laser-induced (multi-photon) trident pair production \cite{nuclearrecoil,nuclearnonrecoil,HCC} and Compton scattering \cite{nonlinear Compton} as well as laser-assisted bremsstrahlung emission \cite{bremsstrahlung}. The cornerstone of the treatment is the laser-dressed QED theory, which by using Volkov states \cite{Volkov}---the eigenstates of a charged particle in the laser field---has taken into account the full orders of the particle-laser interaction, and the remaining weak interaction between the laser-dressed particle and the QED vacuum can be treated perturbatively in a formalism similar to the conventional QED. Although Volkov states play a pivotal role in the theory, as the analytical solution of Dirac equation they can only be obtained for very restricted field geometries among which the most widely used of is the plane-wave. Theoreticians are currently spending much efforts to describe QED effects in short laser pulses \cite{shortlaser} and two-color laser fields \cite{twocolor}, all assuming plane-wave fields. For commonly encountered cases such as colliding lasers and counter-propagating lasers, analytical solution of Dirac equation is not available yet, that cumbers complete QED calculation for those laser configurations.

However, it is worthy to notice that under special conditions laser-dressed QED method based on plane-wave Volkov states can be used to efficiently explore problems in non-plane-wave laser fields, as discussed below.
In this paper complete QED calculation is performed with special focus on the influence of the variation of laser angles on the pair yield and the products' energy spectra to illustrate novel features caused by non-plane-wave nature compared to the plane-wave case. In the following, relativistic units $\hbar=c=\epsilon_0=1$ and a metric such that $a\cdot b=a^\mu b_\mu=a_0b_0-\vec{a}\cdot\vec{b}$ are used.

\begin{figure}
\includegraphics[height=3.2cm,width=5.5cm]{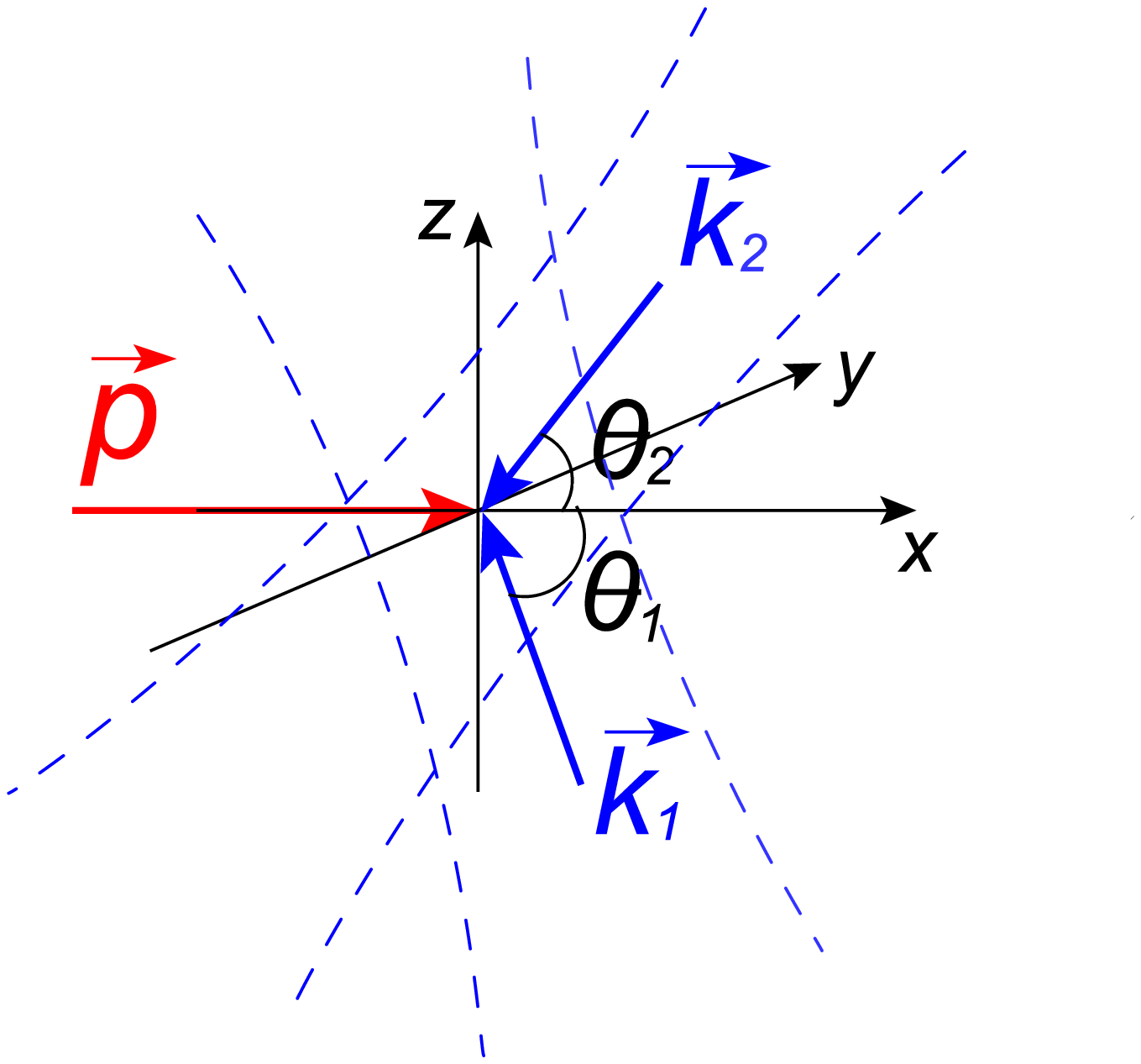}
\caption{\label{coordinate}Coordinates of the collision setup. The dashed lines denote the outlines of the laser beams.}
\end{figure}

The collision setup coordinates are manifested in Fig.\,\ref{coordinate}. The crossing time of the projectile through the beams' overlap region is set to be $\tau\sim40$fs in direct analogy with the SLAC experiment \cite{SLAC, HCC}. This corresponds to the beams' overlap size as well as the respective focal waist size of each beam in the order of  $\sim10\mu$m, which justifies the plane-wave assumption of each beam since the X-ray wavelength $\lambda_{X}=0.12$nm$\ll10\mu$m. Therefore, suppose both X-ray lasers are linearly-polarized plane-wave fields with four-vector potentials $A_i(x)=\epsilon_i a_i \cos(k_i\cdot x)$, $i=1,2$ with $\epsilon_1=\epsilon_2=(0,0,1,0)$, the incoming electron's four-momentum is $p$ and for simplicity $\vec{k}_1$, $\vec{k}_2$ and $\vec{p}$ are coplanar in the $xz$ plane.

Although the absolute intensity of the X-ray lasers here is remarkable, the nondimensional intensity parameter $\xi_{1,2}=e\bar{A}_{1,2}/m\sim0.001$, where $e$ is electron absolute charge, $\bar{A}_{1,2}$ is the root-mean-square value of the laser vector potential, and $m$ is electron mass. Therefore, the laser-particle interaction is in the perturbative regime and the laser modification of electron(positron) mass can be neglected. The energy-momentum criteria $N\omega'\geq 4 m$ with $\omega'$ the energy of the laser photon in the rest frame of the incoming electron, demands that at least two photons must participate in the trident process, while higher order photon absorption processes are suppressed perturbatively.

\begin{figure}
\includegraphics[height=2.5cm,width=9cm]{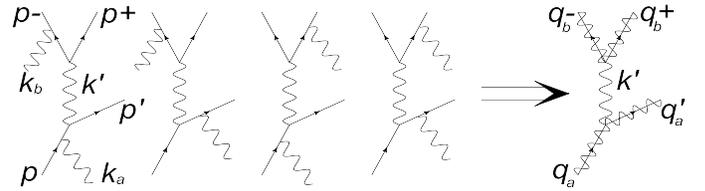}
\caption{\label{feynmanonshell} Left: the 4 types of Feynman diagrams with $k'^2=0$ being possible. They correspond to 16 diagrams as both $k_a$ and $k_b$ have two choices. There are another 16 diagrams due to the exchange of the final indistinguishable electrons.
The complete type set of Feynman diagrams also include those with two laser photons addressed to leptonic lines on one side of the intermediate photon.
Right: the type of Furry-Feynman diagram which can take over the left 4 types of conventional Feynman diagrams. It corresponds to 4 diagrams due to the choices of $k_a$ and $k_b$. There are another 4 diagrams due to the exchange of final electrons.}
\end{figure}

In conventional QED method, the two-photon trident process corresponds to 80 Feynman diagrams, including 20 diagrams for each $e+2k_i\rightarrow e'+e^+e^-$ with $i=1,2$, and the other 40 diagrams for $e+k_1+k_2\rightarrow e'+e^+e^-$. The energy-momentum conservation law allows that for Feynman diagrams of the type specified in Fig.\,\ref{feynmanonshell}, there is a possibility for the intermediate photon to be on-shell. It means that under suitable initial conditions, there exists a range in the momentum space of the scattered electron $p'$ to satisfy  $k'^2=(p-p'+k_i)^2=0$ with $i=1,2$ and $k'$ the momentum of the intermediate photon.
It has been manifested in multi-photon trident pair production in plane-wave lasers that such resonance case corresponds to orders of magnitude of enhancement in pair yield compared to the non-resonance case \cite{HCC,thesis}, similar to the sharp peak in resonance transition in atomic physics \cite{resonanceatom}.
Therefore, in this paper we would only consider the resonance case which accounts for 32 diagrams as illustrated in Fig.\,\ref{feynmanonshell}.

Laser-dressed QED method could be applied to reduce the tedious calculation of the 32 diagrams to 8 Furry-Feynman diagrams where particles are represented by Volkov states instead of free-field states. It is illustrated \cite{thesis,HC} that at the limit $\xi\rightarrow 0$, up to a normalization factor of the field, the proper amplitude constituent of a vertex with one photon attached to one of its leptonic legs can be restored from the first-order Fourier expansion term of the amplitude expression of a laser-dressed vertex. By assigning different dressing lasers at the two vertices, the amplitude of trident pair production in two colliding laser beams can be obtained, as long as the condition $\xi\ll1$ is fulfilled. Close study reveals that the resonance condition further renders the total amplitude square calculation to the sum of amplitude squares of the 4 Furry-Feynman diagrams shown in Fig.\,\ref{feynmanonshell}.  The interference terms caused by the exchange of final electrons can be dropped, because the two diagrams with exchanged final electrons can seldom satisfy the resonance condition simultaneously, leading to negligible interference.

\begin{figure}
\includegraphics[height=4.2cm,width=6.5cm]{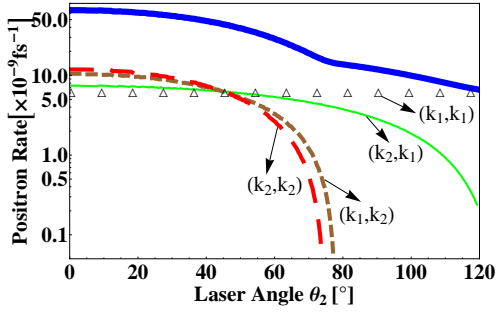}
\caption{\label{vak2} The total rate (blue thick line) and the respective contributions from the four channels (marked by channel names) for $\theta_1=-45^\circ$ and variant $\theta_2$. }
\end{figure}

The total amplitude square reads
\begin{align}\label{smatrix0}
|S_{fi}|^2=\sum_{u=1,2}\sum_{v=1,2}|\mathcal{M}(q_u,q'_u,q^+_v,q^-_v)|^2\,,
\end{align}
where the subscript $u,\,v$ indicates whether the particle is dressed by laser $A_1$ or laser $A_2$, and
\begin{align}\label{smatrix1}
\mathcal{M}(q_u,q'_u,&q_v^+,q_v^-)=-i\alpha\int d^4x\int d^4y \bar{\Psi}_{q'_u}(x) \gamma^\mu \Psi_{q_u}(x)\nonumber\\
&\times D_{\mu\nu}(x-y)\bar{\Psi}_{q^-_v}(y) \gamma^\nu \Psi_{q^+_v}(y)\,,
\end{align}
where $\alpha$ is the fine-structure constant, $D_{\mu\nu}(x-y)=\int\frac{d^4k'}{(2\pi)^4}\frac{-i g^{\mu\nu}}{k'^2+i\epsilon}e^{ik'\cdot (x-y)}$ is the free photon propagator, and $\Psi_{q_u}, \Psi_{q'_u}, \Psi_{q^+_v}, \Psi_{q^-_v}$ denote the laser-dressed lepton states, which can be decomposed into sums of Fourier series of $e^{ink_{u}x}$ or $e^{imk_{v}y}$. Performing the space-time integration of Eq. (\ref{smatrix1}) gives
\begin{align}\label{M_fi}
&\mathcal{M}(q_u,q'_u,q_v^+,q_v^-)=\beta(q_u,q'_u,q_v^+,q_v^-)\displaystyle\sum_{n}\sum_{m} M^\mu(q_u,q'_u|n) \\
&\times M_\mu(q_v^+,q_v^-|m) \nonumber\frac{\delta^{(4)}(q_u+nk_u+mk_v-q'_u-q_v^+-q_v^-)}{(q_u-q'_u+nk_u)^2+i\epsilon}\, ,
\end{align}
where $n,\,m$ correspond to the number of photons absorbed by the leptonic lines connected to the two vertices, respectively, and $n=m=1$ for the diagrams considered here.
As $\xi\ll1$, to the lowest order in $A_1$ and $A_2$, the terms in Eq. (\ref{M_fi}) turn out to be
\begin{align}\label{spinor}
 M^\mu(q_u,&q'_u|1)=\bar{u}_{p's'}\{\frac{1}{2}(\frac{ea_u (\epsilon_u\cdot p')}{k_u\cdot p'}-\frac{ea_u(\epsilon_u \cdot p)}{k_u \cdot p})\gamma^\mu \nonumber\\
& -\frac{1}{2}(\frac{ea_u\slashed{\epsilon}_u\slashed{k}_u\gamma^\mu}{2k_u\cdot p'}
+\frac{ea_u\gamma^\mu\slashed{k}_u\slashed{\epsilon}_u}{2k_u \cdot p})\}u_{p,s}\,,\nonumber\\
 M_\mu(q_v^+&,q_v^-|1)=\bar{u}_{p_-s_-}\{\frac{1}{2}(\frac{ea_v(\epsilon_v\cdot p_-)}{k_v\cdot p_-}-\frac{ea_v (\epsilon_v\cdot p_+)}{k_v\cdot p_+})\gamma_\mu\nonumber\\
& -\frac{1}{2}(\frac{ea_v\slashed{\epsilon}_v\slashed{k}_v\gamma_\mu}{2k_v\cdot p_-}
-\frac{ea_v\gamma_\mu\slashed{k}_v\slashed{\epsilon}_v}{2k_v\cdot p_+})\}u_{p_+s_+}\,.
\end{align}
After substituting Eqs. (\ref{spinor}) into Eq. (\ref{M_fi}), only the coefficient $\beta$ and the fraction containing the $\delta$ function are left with the laser-dressed four-momentum, e.g., $q_u$, which is related to the free four-momentum $p$ outside the field by $q_u=p+\frac{m^2 \xi_u^2}{2 k_u\cdot p}k_u$ with $\xi_u=e\bar{A}_u/m\,,u=1,2$. As discussed above, $\xi_u\ll1$ and therefore $q_u\approx p$, that in the subsequent calculation all dressed momenta are substituted by their corresponding free momenta  and the coefficient $\beta=-2\alpha (2\pi)^5(\frac{m^4}{p^0p'^0p^{+0}p^{-0}V^4})^\frac{1}{2}$ with $V$ being the computational volume.

The total rate is obtained as
\begin{equation}\label{rate}
R = \frac{1}{T} \int\frac{d^3p^+}{(2\pi)^3} \int\frac{d^3p^-}{(2\pi)^3} \int\frac{d^3p'}{(2\pi)^3} \frac{1}{2}\sum_{\rm spins} |S_{fi}|^2\,
\end{equation}
with the interaction time $T$ and a statistical factor $1/2$ due to initial spin averaging. The spin sum is converted in the usual way into voluminous trace products. The resonance condition indicates that there are denominators like $(q_u-q'_u+nk_u)^2+i\epsilon\approx(p-p'+nk_u)^2+i\epsilon=0+i\epsilon$ in $S_{fi}$, which result in poles in the integration of Eq. (\ref{rate}) and require special mathematical treatments. Detailed derivation and procedure of calculation are manifested in \cite{thesis}. The basic idea is to regularize the pole by the finite interaction time of the process. The two main competing mechanisms in the case are decaying of the Volkov states due to Compton scattering and decaying due to the particles passing through the boundaries of the laser beams, between which the shortest lifetime defines the interaction time. Here the decaying due to Compton scattering is orders of magnitude slower than due to the passing boundary mechanism. Therefore $\tau\sim40$fs is adopted as the regularization time. Finally a numerical Monte Carlo approach is employed for the multi-dimensional integration.

\begin{figure}
\includegraphics[height=3.5cm,width=9cm]{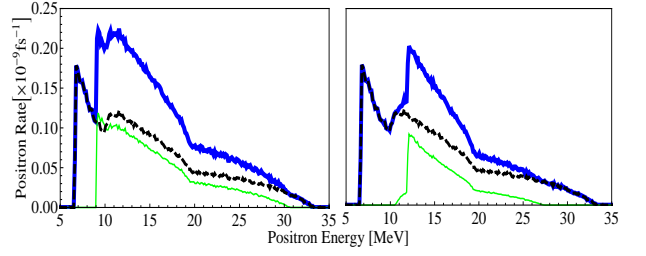}
\caption{\label{momentum} The total positron energy distribution (blue thick line) and the respective contributions from $(k_1,k_1)$ channel (black dashed line) and $(k_2,k_1)$ channel (green thin line) for $\theta_1=-45^\circ$ and (left) $\theta_2=80^\circ$, (right) $\theta_2=100^\circ$.}
\end{figure}

To clarify the different contributions of the two beams, the Furry-Feynman diagram in Fig.\,\ref{feynmanonshell} can be seen as representing four channels: channel $(k_1,k_1)$ is where both photons of $k_a$ and $k_b$ are from laser $A_1$, channel $(k_2,k_2)$ is where both photons are from $A_2$, channel $(k_1,k_2)$ is where $k_a$ is $k_1$ and $k_b$ is $k_2$, and channel $(k_2,k_1)$ is vice verse. For fixed $A_1$ beam with $\theta_1=-45^\circ$, the total rate as well as the respective contributions from the four channels is investigated with the variation of $\theta_2$, shown in Fig.\,\ref{vak2}. Except for the channel $(k_1,k_1)$ which depends only on the fixed beam $A_1$, the rate of the channel $(k_2,k_1)$ drops notably slower than the other two channels with the increase of $\theta_2$. In the range $80^\circ\leq\theta_2\leq120^\circ$, the total rate can be seen as composed only of the rate of channel $(k_2,k_1)$ and a constant baseline rate of $(k_1,k_1)$. In this sense, the channel $(k_2,k_1)$ can be distinguished and studied separately. Note that the laser intensity dependence of the total rate can not be simply interpreted by the conventional perturbation theory that the two-photon absorption rate is proportional to the square of the laser field intensity.

The angle effect can also be identified in the positron energy distribution, shown in Fig.\,\ref{momentum}, where the laser beam setup makes the contributions from channel $(k_1,k_2)$ and $(k_2,k_2)$ negligible, as discussed above. Different from the single-peak structure of the positron energy distribution found in the SLAC experiment \cite{SLAC}, there are generally more than one peak in the colliding beam case. In Fig.\,\ref{momentum}, the first peak is solely settled by channel $(k_1,k_1)$, and the position and shape of the second peak is largely determined by channel $(k_2,k_1)$. The onset of the second peak changes notably from 9MeV to 12MeV as $\theta_2$ turns from 80$^\circ$ to 100$^\circ$. Therefore, the positron energy measurement provides a good test and sensitive indicator for studying the colliding beam angle effect.

\begin{figure}
\includegraphics[height=3.5cm,width=9cm]{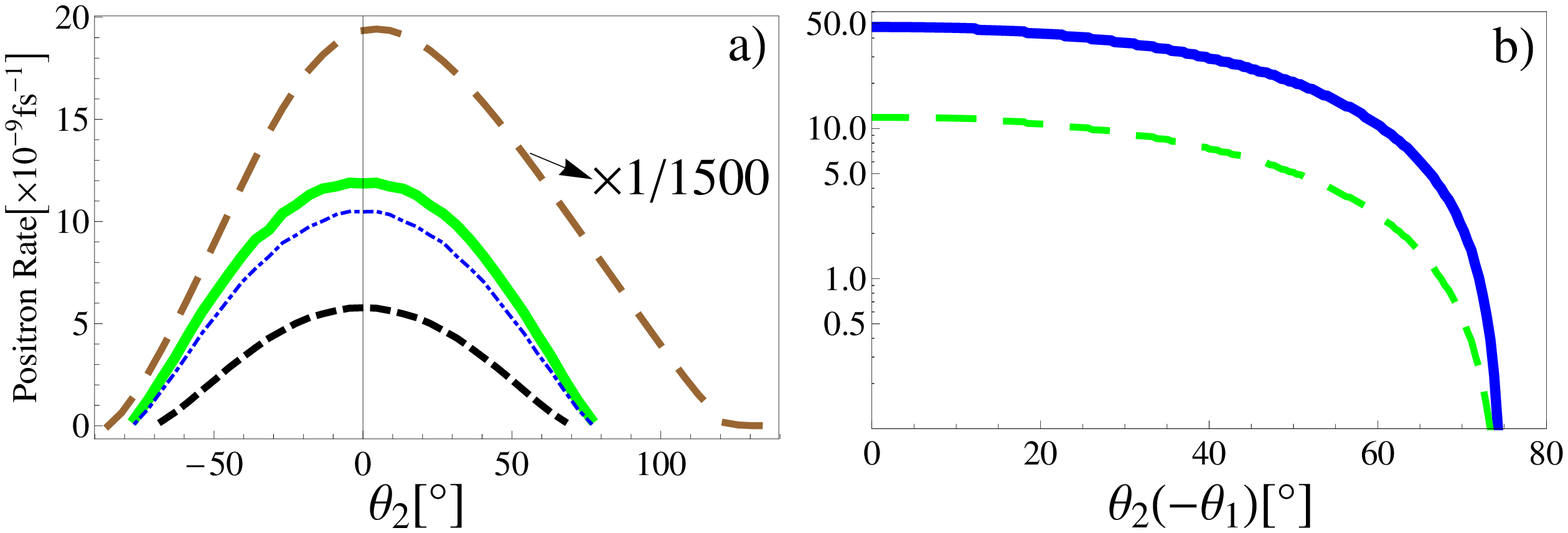}
\caption{\label{useonetwo}a) Positron production rate of channel $(k_1,k_2)$ is plotted against $\theta_2$ for $\theta_1=0^\circ$ (green solid line), $\theta_1=-45^\circ$ (blue dash-dot line) and $\theta_1=-90^\circ$ (black dashed line at the bottom), all of which present the numerical symmetry about $\theta_2=0^\circ$. The top brown dashed line manifests the reduced (by a factor of $1/1500$) rate of channel $(k_1,k_2)$ for incoming electron energy $1.2$MeV, laser frequency $0.5$MeV and $\theta_1=-90^\circ$ with otherwise the same parameters as the rest of the paper. b) Angle dependence of positron production rate of each single channel (green dashed line) and the total (blue solid line) in symmetric configurations ($\theta_2=-\theta_1$).}
\end{figure}

Fig.\,\ref{vak2} is also featured that the respective rates of the four channels coincide at $\theta_2=-\theta_1=45^\circ$. This is not merely a coincidence. As manifested in Fig.\,\ref{useonetwo}a) where channel $(k_1,k_2)$ is calculated with variant $\theta_2$ for different fixed $\theta_1$, the rate curve is nearly symmetric about $\theta_2=0^\circ$. Since the setup with $\theta_1$ and $\theta_2$ is essentially equivalent to that with $-\theta_1$ and $-\theta_2$ which gives nearly the same rate as that with $-\theta_1$ and $\theta_2$,  the rate should also be symmetric about $\theta_1=0^\circ$ for different fixed $\theta_2$. Similar argument also applies for channel $(k_2,k_1)$. As the rate of channel $(k_i,k_i)$ with $i=1,2$ only depends on $|\theta_i|$, therefore the configurations with $(\theta_1,\theta_2)$, $(-\theta_1,\theta_2)$, $(\theta_1,-\theta_2)$ and $(-\theta_1,-\theta_2)$ can not be distinguished by comparing their production rates.
It should be emphasized that the angle symmetry found here is tenable only numerically. The cause of the symmetry is attributed to the high energy of the incoming electron that renders the momentum direction of the intermediate photon distributed nearly along the incoming electron momentum direction, that $0^\circ$. Thus the subsequent collision of the intermediate photon with the second laser photon $k_b$ is insensitive to the sign of $\theta_b$. For comparable incoming electron energy and laser photon energy, this symmetry does not maintain as displayed in Fig.\,\ref{useonetwo}a).

Due to the numerical symmetry, it can be inferred that for symmetric configurations as $\theta_2=-\theta_1$,
despite the variation of the laser angle, almost the same production rates are given by the four channels. Calculated results are illustrated in Fig.\,\ref{useonetwo}b). The two laser beams increase the total rate to four times of that by a single laser beam, and thus the conventional perturbation explanation of the production rate dependence on the laser field intensity applies here.

It can be seen from Fig.\,\ref{vak2} and Fig.\,\ref{useonetwo} that for the designated laser and projectile parameters, the total positron production rate per projectile can reach $\approx5\times10^7$s$^{-1}$, corresponding to the head-on collision $\theta_1=\theta_2=0^\circ$, and for a wide range of variant laser angles, the rate can maintain well above $10^6$s$^{-1}$. This is 2$\sim$3 orders of magnitude higher than that found in the SLAC experiment \cite{SLAC,HCC}. Thus the realization of experimental observation is promising.

A different laser polarization condition is also calculated for $\vec{\epsilon}_1$ and $\vec{\epsilon}_2$ in the $xz$ plane, e.g., $\epsilon_1=(0,k_1^3/k_1^0,0,-k_1^1/k_1^0)$ and $\epsilon_2=(0,k_2^3/k_2^0,0,-k_2^1/k_2^0)$. Negligible difference has been found in the pair yield compared with the previous polarization case. This indicates that in the high energy collision involving high frequency lasers with relatively small field parameter $\xi$, the role played by the laser field is mainly determined by the photon momentum, while the laser polarization could be important in other scenarios such as when $\xi$ being large.

In summary, a complete laser-dressed QED calculation for a resonant two-photon trident electron-positron pair production in the interaction of an energetic electron with two colliding bright X-ray laser beams has been performed. The total as well as channel specified pair production rate and positron energy spectrum are explored, focusing on the angle effects. Novel features due to the non-plane-wave colliding laser configuration have been identified. The influence of relative phase and dissimilar frequencies between the two laser beams at the overlap region can be investigated in further studies.

We are grateful to Prof. C. M\"uller for fruitful discussions and valuable advices. This work is financially supported by the National Natural Science Foundation of China under Grant No. 11204370.

\end{document}